\documentclass[journal,twoside,web]{ieeecolor}
\usepackage{lcsys}
\usepackage{cite}
\usepackage{amsmath,amssymb,amsfonts}
\usepackage{algorithmic}
\usepackage{graphicx}
\usepackage{textcomp}
\def\BibTeX{{\rm B\kern-.05em{\sc i\kern-.025em b}\kern-.08em
    T\kern-.1667em\lower.7ex\hbox{E}\kern-.125emX}}
\usepackage{todonotes}
\usepackage[ruled,vlined]{algorithm2e}
\usepackage{mdframed}
\usepackage{soul}
\usepackage{amsfonts}

\newcommand{\R}{\mathbb{R}}
\newcommand{\0}{\mathsf{0}}

\sethlcolor{blue!30}

\usepackage{amsmath, amssymb}
\usepackage{tikz}

\newtheorem{definition}{Definition}
\newtheorem{proposition}{Proposition}

\usetikzlibrary{positioning, arrows}

\begin{document}
\title{Fault-Tolerant MPC Control for Trajectory Tracking}
\author{David Laranjinho \and Daniel Silvestre
	\thanks{David Laranjinho and Daniel Silvestre are with the School of Science and Technology, NOVA University of Lisbon, 2829-516 Caparica, Portugal, also with CTS - Centro de Tecnologia e Sistemas part of LASI, 2829-516 Caparica. D. Silvestre is also with the Institute for Systems and Robotics, Instituto Superior Técnico, University of Lisbon, 1049-001 Lisbon, Portugal (e-mail: dsilvestre@fct.unl.pt).}
	\thanks{The work from D. Silvestre was funded by FCT – Fundação para a Ciência e a Tecnologia through CTS - Centro de Tecnologia e Sistemas/UNINOVA/FCT/NOVA with reference UID/0066/2025.}
}

\maketitle

\begin{abstract}
An MPC controller uses a model of the dynamical system to plan an optimal control strategy for a finite horizon, which makes its performance intrinsically tied to the quality of the model. When faults occur, the compromised model will degrade the performance of the MPC with this impact being dependent on the designed cost function. In this paper, we aim to devise a strategy that combines active fault identification while driving the system towards the desired trajectory. The explored approaches make use of an exact formulation of the problem in terms of set-based propagation resorting to Constrained Convex Generators (CCGs) and a suboptimal version that resorts to the SVD decomposition to achieve the active fault isolation in order to adapt the model in runtime.
\end{abstract}

\begin{IEEEkeywords}
Fault-tolerant Control, reachability analysis, Singular Value Decomposition.
\end{IEEEkeywords}

\section{Introduction}
The MPC controller is an optimal control strategy that solves an optimization problem for a given horizon with a known model for the system. If the model is incorrect, the mismatch between the model and the plant will degrade the performance as the prediction in the horizon will not hold in practice. In the presence of faults, the control strategy should detect the model change while still attempting to drive towards the desired trajectory point. Within the field of the Aerospace industry, many applications can be described by nonlinear models with the nonlinear terms involving parameters like the angular velocity or inertia and driven by thrusters that are characterized by a positive input \cite{taborda:mpc}. With that in mind, we will focus on Fault-tolerant MPC applied to uncertain Linear Parameter Varying (LPV) systems where the control action is positive, which can model from quadrotors to spacecraft. \par

A related concept to the considered setup in this paper is the theory of positive systems where states and outputs remain non-negative for all time whenever the initial conditions and inputs are non-negative. In the linear case, this property is guaranteed when the dynamics matrix is Metzler with nonnegative input, output, and feed through matrices. Positive systems model processes where negative values have no physical meaning and have simpler stability analysis, performance evaluation, and controller design. For example, stability can often be certified using linear rather than quadratic Lyapunov functions, and optimization-based methods such as model predictive control benefit from reduced computational complexity. In this paper, we consider only that the control action is non-negative with no further assumptions. \par

There are two different approaches when it comes to the design of Fault Detection and Isolation (FDI): i) model-based and ii) data-driven. Beginning with model-based FDI, first introduced in 1971 and studied ever since \cite{model_based_fdi_1} \cite{model_based_fdi_2} \cite{model_based_fdi_3}, this approach is based on estimating the residuals by designing an observer that estimates the output of the plant and monitor its difference against the acquired measurements. We can then find a fault and isolate it by calculating the residue between the output and the estimated output. Examples like constructing a bank of Kalman Filters \cite{fdi_kalman}, $H_{\infty}$ filters in linear \cite{fdi_H_inf_linear} and non linear \cite{fdi_H_inf_non_linear} cases as well as Sliding Mode Observers (SMO) \cite{fdi_smo} have all been considered in the literature. This concept can also be considered when the observers produce sets \cite{fdi_set} to isolate actuator faults \cite{FDI_observers}, or for the case of multiple controllers \cite{FDI_observer_mult_contr} and even in a distributed setting \cite{silvestre:distributedFDI}.

The second approach is data-driven, where it is not required the plant description and FDI is accomplished by analyzing the previous outputs of the system either with regression, an Artificial Intelligence model, time/frequency analysis, Singular Value Decomposition (SVD) \cite{aircraft_fault_detection}, among others. The main focus of these methods is detecting a general trend in the past input-output relation to detect an anomaly in the present input-output relation. A recent method that makes sole use of the input-output relation in order to perform FDI on a system, uses a SVD approach on a matrix that encompasses the time responses arranged into a Henkel matrix, which is equivalent to multiplying the observability and controllability matrices together \cite{fdi_svd}. Given that some faults might not sufficiently excite the output measurements, active FDI improves on this issue by injecting specific inputs to get more information of the system, with the disadvantage of temporarily reducing system performance.

When resorting to set-based FDI, the residual logic is replaced by the idea of separating the reachable sets of the various models. One such approach to LTI systems makes use of the gain of a Luenberger observer to create the separation \cite{LTI_sep_paper}. This introduces some conservatism as it uses an observer that adds some regions through the Minkowski operator to the original set and is only applicable to systems without uncertainty, making it undesirable for nonlinear systems that can be modeled by an LPV.
The authors of \cite{Set-Based-afd} introduced the idea of constructing sets representing all possible combinations of modes and control actions, given bounded uncertainty in a high-dimensional space comprised of the states and control actions throughout the entire horizon. The next step involves extracting the subspace of this set that contains the intersection of the modes. The main disadvantage is the high computational cost given by vertex enumeration or Fourier Motzkin for the exact projection associated with retrieving the control action subspace. In a similar direction, \cite{paper_application1} designs fault-tolerant controllers for systems with LPV models and unknown bounded uncertainty. The paper in \cite{paper_application2} still resorted to the active identification by separating sets while considering Gaussian noise represented by the confidence ellipsoids for use cases with LTI systems. \par

In the literature of set-based estimation, it has recently been introduced the concept of Constrained Convex Generators (CCG) \cite{ccg} arising as a generalization of Constraint Zonotopes (CZ). Since CCGs can represent both polytopes and ellipsoids, they are better suited to model cases with energy constraints arising from physical quantities like velocity (ellipsoids) and use polytopes to represent limit bounds on the sensors. Moreover, it was proved in \cite{silvestre:exactCvxHull} that CCGs admit an optimal representation for the convex hull operation, which will greatly reduce the computational complexity for the broad family of uncertain LPVs. Therefore, the main objective in this paper is to generalize the concept of active FDI in \cite{Set-Based-afd} to handle energy-like constraints and avoid the representation using polytopes in CZ and H-representations. By representing the sets with CCGs, and since these have an optimal representation both in accuracy and data structure size for the convex hull operation (see the exact expression in \cite{silvestre:exactCvxHull} and a result highlighting the growth when using CZs in \cite{silvestre:accurateCvxHull}), we are able to outperform the state-of-the-art both in terms of accuracy and computational complexity. We also present a sub-optimal approach based on the SVD that further reduces the computational time and memory footprint for cases where the exact representation might not be needed.

\par \textbf{Notation}: Throughout this paper, matrices are denoted by capital letters while vectors are lower case letters, with the distiction that $\mathbf{u}$ will be used to represent the control sequence from $k=0$ to $k=N$ while $u_k$ is used to represent one control action at time $k$. The transpose of a matrix $A$ is denoted by $A^\intercal$. We will use the notation $\mathcal{B(\theta)} = \mathbf{Conv} \left( \bigcup_{v \in \mathrm{vertex(\Theta)}} B(v)\right) $ to denote the polytopic set of the matrices generated by the uncertain vector $\theta$ and similarly for the $A$ matrix. The controllability matrix is given by
\begin{equation}
	\mathcal{C}_N = 
	\begin{bmatrix}
		B & A B & A^2 B & \cdots & A^{N-1} B
	\end{bmatrix}
\end{equation}
for matrices $A, B$ representing the dynamics and input matrices. We will also make use of tuples of matrices and vectors for a shorthand notation of sets described by those values like in $\mathcal{Z} = \left( G_z, c_z, A_z, b_z, \xi_z \right)$.

\section{Problem Formulation}

In this paper, we consider a dynamical system that can be described by the model
\begin{equation}
\label{eq:lpv}
\begin{aligned}
x_{k+1} &= A(\theta_k)x_k + B(\theta_k)u_k + r(\theta_k) + E(\theta_k)w_k, \\
y_k &= C(\theta_k)x_k + s(\theta_k) + F(\theta_k)v_k. 
\end{aligned}
\end{equation}
where $x \in \mathbb{R}^{n_x}$, $u \in \mathbb{R}^{n_u}_+$ and $y \in  \mathbb{R}^{n_y}$ represent the state, input, and output signal of the system, respectively. Additionally, $w \in  \mathbb{R}^{n_w}$ is the disturbance of the process and $v \in  \mathbb{R}^{n_v}$ the measurement noise. The values $r$ and $s$ are additive bias to the actuator and sensor, whereas $\theta_k$ contains the values of the parameters for the LPV at time $k$. 

In order to formulate the problem of active FDI using a set-based approach, we require a formal definition of distinguishability for linear systems that was given in ~\cite{LOU200921}. \par
\textbf{Definition (Input Distinguishability):}  \textit{Given} $T > 0$, \textit{let} $\mathcal{U}$ be a set describing all possible values of $\mathbf{u}$ for the horizon $T$. \textit{We say that systems } $S_1$ \textit{and} $S_2$ \textit{with models following \eqref{eq:lpv} are \textbf{input distinguishable} on} $[0, T]$, \textit{if, for any non-zero}
\begin{equation}
	( x_0^{(1)} , x_0^{(2)} , \mathbf{u}) \in \mathbb{R}^n \times \mathbb{R}^n \times \mathcal{U}
\end{equation}
\textit{the outputs $y^{(1)}_k$ and $y^{(2)}_k$ can not be identical to each other on $[0, T]$ regardless of the values of the initial conditions $x_0^{(1)} , x_0^{(2)}$, respectively for systems }$S_1$ \textit{and} $S_2$.

Notice that even though the input distinguishability is theoretical, we can construct a set encoding all possible values for the signals for which the two systems are distinguishable as was done in \cite{distinguishing_discrete_time_linear_sys} using the concept of Set-Valued Observers (SVOs) that consists in storing this relationship as a polytope with a vector containing all the signals. Following the concept of distinguishability, fault detection corresponds to checking if the reachable sets for the nominal and potential faulty models do not intersect. This would mean that \( x_k \) violates the reachable set for the nominal model computed from prior information, i.e., 
\begin{equation}
	x_k \notin \mathcal{X}_k \Rightarrow \text{Fault Detected}.
\end{equation}

In a similar fashion, fault isolation extends this concept to multiple sets by having a bank of observers each for a particular faulty model. Therefore, the problem addressed in this paper is the design of the actuation signal $u_{k}$ such that the sets for the faulty and nominal models do not intersect while attempting to track a reference. For simplicity of exposition, we will be denoting $\tilde{x}^{(1)}_k = x^{(1)}_k - \bar{x}$ for some reference $\bar{x}$. Therefore, the optimization problem being tackled corresponds to
\[
\begin{aligned}
\min_{\{u_{k+i}\}_{i=0}^{N-1}} \;& \sum_{i=0}^{N-1} \big( \tilde{x}_{k+i}^{(1)\intercal} Q \tilde{x}_{k+i}^{(1)} + u_{k+i}^\intercal R_1 u_{k+i}  + \\
& \|R_2 (u_{k+i+1} - u_{k+i})\|^2 \big) 
+ \tilde{x}_{k+N}^{(1)\intercal} P \tilde{x}_{k+N}^{(1)} \\
\text{s.t.}\;& x_{k+i+1}^{(1)} \text{ satisfies \eqref{eq:lpv}}, \quad i = 0,\dots,N-1, \\
& x_{k+N}^{(1)} \in \mathcal{X}_f^{(1)}, \quad u_{k+i} \in \mathcal{U}, \quad i = 0,\dots,N-1, \\
& \mathcal{X}_{k+N}^{(1)} \cap \mathcal{X}_{k+N}^{(2)} = \emptyset,
\end{aligned}
\]
where the sets $\mathcal{X}_{k+N}^{(1)}$ and $\mathcal{X}_{k+N}^{(2)}$ are built for the nominal and faulty model in consideration, respectively. As soon as a fault is detected (i.e., $\mathcal{X}_{k}^{(1)}$ is empty for some time instant $k$), the new model being considered with be the faulty one and the same problem stated above can be solved with different models for a more accurate isolation or detect recovery.

\section{Optimal Set Separation with CCG}
\label{sec:sep_ccg}

Before introducing the solution to the active FDI, we recover the definition of a CCG and also the main operations with closed-form expression that can be found in \cite{ccg}.

\begin{definition}[Constrained Convex Generators]
	\label{def:CCG}
	A Constrained Convex Generator (CCG) $\mathcal{Z} \subset \R^{n}$ is defined by the tuple $(G,c,A,b,\mathfrak{C})$ with $G \in \R^{n\times n_{g}}$, $c \in \R^{n}$, $A \in \R^{n_{c}\times n_{g}}$, $b \in \R^{n_{c}}$, and $\mathfrak{C} := \{\mathcal{C}_{1}, \mathcal{C}_{2}, \cdots, \mathcal{C}_{n_{p}}\}$ such that:
	\begin{equation}
		\mathcal{Z} = \{G \xi + c :  A\xi = b, \xi \in \mathcal{C}_{1} \times \cdots \times \mathcal{C}_{n_{p}}\}.
	\end{equation}
\end{definition}

\begin{proposition}
	\label{pro:CCG}
	Consider three Constrained Convex Generators (CCGs) as in Definition \ref{def:CCG}:
	\begin{itemize}
		\item $\mathcal{Z} = (G_{z}, c_{z}, A_{z}, b_{z},\mathfrak{C}_{z}) \subset \R^{n}$;
		\item $\mathcal{W} = (G_{w}, c_{w}, A_{w}, b_{w},\mathfrak{C}_{w}) \subset \R^{n}$;
		\item $\mathcal{Y} = (G_{y}, c_{y}, A_{y}, b_{y},\mathfrak{C}_{y}) \subset \R^{m}$;
	\end{itemize}
	and a matrix $R \in \R^{m \times n}$ and a vector $t \in \R^{m}$. The three set operations are defined as:
	\begin{equation}
		\label{eq:CCGlinearmap}
		R\mathcal{Z}+t = \left(RG_{z}, Rc_{z}+t, A_{z},b_{z},\mathfrak{C}_{z}\right)
	\end{equation}
	\begin{equation}
		\label{eq:CCGminkowski}
		\mathcal{Z} \oplus \mathcal{W} = \left(\begin{bmatrix}
			G_{z} & G_{w}
		\end{bmatrix}, c_{z} + c_{w}, \begin{bmatrix}
			A_{z} & \0 \\
			\0 & A_{w}
		\end{bmatrix},\begin{bmatrix}
			b_{z} \\ b_{w}
		\end{bmatrix},\{\mathfrak{C}_{z},\mathfrak{C}_{w}\}\right)
	\end{equation}
	\begin{equation}
		\label{eq:CCGintersection}
		\resizebox{\hsize}{!}{$
		\mathcal{Z} \cap_{R} \mathcal{Y} = \left(\begin{bmatrix}
			G_{z} & \0
		\end{bmatrix}, c_{z}, \begin{bmatrix}
			A_{z} & \0 \\
			\0 & A_{y} \\
			RG_{z} & -G_{y}
		\end{bmatrix},\begin{bmatrix}
			b_{z} \\ b_{y} \\ c_{y} - Rc_{z}
		\end{bmatrix},\{\mathfrak{C}_{z},\mathfrak{C}_{y}\}\right).$}
	\end{equation}
\end{proposition}

Since the optimization associated with the MPC formulation requires building the reachable sets based on available dynamical equations, we start by considering the definition of reachable set

\begin{equation}
\label{eq:reach}
\begin{aligned}
\mathcal{R}_N
=
\Big\{&
[y_{k+N},\, u_k,\, \ldots,\, u_{k+N-1}]
\;\Big|\;
\Big.
\\[0.5ex]
\Big.
& x_{k+i+1} = A(\theta_{k+i}) x_{k+i} + B(\theta_{k+i}) u_{k+i} + r(\theta_{k+i}) + \\ 
&E(\theta_{k+i}) w_{k+i}, \\
& y_{k+i} = C(\theta_{k+i})x_{k+i} + s(\theta_{k+i}) + F(\theta_{k+i}) v_{k+i} \\
& x_0 \in \mathcal{X}_0,\;
u_{k+i} \in \mathcal{U},\;
w_{k+i} \in \mathcal{W}, \;
v_{k+i} \in \mathcal{V}\\
& i = 0,\dots,N-1\Big\}.
\end{aligned}
\end{equation}

From the definition of reachable set, we can take the scheduling vector $\theta_k$ to be equal to a vertex of the set of dynamics matrices, apply the formulas in Proposition \ref{pro:CCG} to replace vector operations with that of sets, and then perform the convex hull over all sets produced for each vertex. Since the typical MPC problem minimizes a quadratic cost function corresponding to the system energy, it is sufficient to consider the boundary of the indistinguishable region, rather than computing the full complement. This significantly improves computational efficiency as the problem reduces to finding the boundary of a set in $\mathbb{R}^{n_u N}$, rather than enumerating vertices in the full $(n_x+n_u) N$ space, as in \cite{Set-Based-afd}. Moreover, as we have seen from the definition of set operations for CCGs, these have closed-form expressions for all the steps required to build the reachable set. Therefore, the proposed technique in this paper is to build the reachable sets for the LPV model and enforce its separation with the same set with a fault present. Our approach is summarized in Algorithm \ref{alg:ccg_sep}.

\begin{algorithm}
\label{alg:ccg_sep}
\DontPrintSemicolon
\KwIn{Initial set $\mathcal{X}_0$, noise set $\mathcal{V}$, disturbance set $\mathcal{W}$, scheduling polytope $\Theta$, and feasible input set $\mathcal{U}$, weight matrix for control action energy $R_1$ and variation $R_2$, list of modes ${M}$, regulation factor $\gamma$}
\KwOut{control sequence for FDI $u^\star$}
\BlankLine
Initialize reachable set collection $\mathcal{R} \gets \emptyset$\;
Initialize indistinguishable set collection $\mathcal{I} \gets \emptyset$\;
Compute the subsets of carnality 2 $P_2 \gets \binom{M}{2}$ \;
\ForEach{m $\in M$}{
\ForEach{$\theta \in \Theta$}{
Compute reachable set $\mathcal{R}_\theta^m$ in extended state-control space\;
$\mathcal{R}^m \gets$ Convex Hull($\mathcal{R}^m, \mathcal{R}_\theta^m$);}}
\ForEach{${i,j} \in P_2$}{
$\mathcal{I}^{i,j} \gets$ Intersection($\mathcal{R}^i, \mathcal{R}^j$)\;
$\mathcal{I}^{i,j} \gets$ Project $\mathcal{I}^{i,j}$ onto sub-space $\mathcal{U}$\;
$\mathcal{I} \gets$ Convex Hull($\mathcal{I}, \mathcal{I}^{i,j}$)}\;
$u^\star \gets \arg\min_{u \in \mathcal{U} \setminus \mathcal{I}}
\;\gamma\, \mathbf{u}^\intercal R_1 \mathbf{u} + (1-\gamma)\,\lVert R_2 \mathbf{u} \rVert_2^2$\;

\caption{Optimal Separation with CCG}
\end{algorithm}

From the pseudo-code in Algorithm \ref{alg:ccg_sep}, it becomes clear that two additional operations are required for the proposed FDI sets, namely i) the product of sets required for the product between the polytopic set of the input matrix and the set describing all possible control actions; ii) the computations in state-control space. In the remainder of this section, we provide the details for these two operations.

\subsection{Multiplication of Sets}
In the construction of the reachable set, we will need the product between the set resulting from the convex hull of the controllability matrices under any scheduling vector value with the set of feasible inputs. In mathematical terms, this operation arises from the product $\mathcal{B}(\theta) \mathcal{U}$ where both $\mathcal{B}(\theta)$ and $\mathcal{U}$ are sets. In this paper, we adopt a construction similar to the convex hull operation for CCGs and handle each entry $\alpha$ following
\begin{align}
  \mathcal{B}(\theta)\mathcal{U}_\alpha &= (G_B \xi_B + c_B) u_\alpha \nonumber \\
            &= G_B \xi_B \, u_\alpha + c_B (G_{u_\alpha} \xi_u + c_{u_\alpha}) \nonumber \\
            &= G_B \xi_B G_{u_\alpha} \xi_{u_\alpha} + G_B \xi_B c_{u_\alpha} + c_B G_u \xi_u + c_B c_{u_\alpha}
\end{align}

The first term of the expression is nonlinear, as it involves the product of two generators. Consequently, it cannot be implemented directly within the canonical CCG format. However, by introducing norm cone constraints this is made possible. More precisely, instead of computing the bilinear term explicitly, we introduce an auxiliary generator that bounds the interaction term. By making $||\xi_B|| \leq G_{u_\alpha} \xi_{u_\alpha} + c_{u_\alpha}$ we have made it so the generator is proportional to the amount we want to multiply, leading to
\begin{equation}
  \begin{bmatrix}
    x \\
    u
  \end{bmatrix}
  =
  \begin{bmatrix}
    G_B & G_{u_\alpha} c_B \\ 0 & G_{u_\alpha} 
  \end{bmatrix}
  \begin{bmatrix}
  \xi_B \\
  \xi_{u_\alpha}
  \end{bmatrix}
  +   \begin{bmatrix}
  c_{u_\alpha} c_B \\
  c_{u_\alpha}
  \end{bmatrix}
\end{equation}

\begin{equation}
\label{eq:conicGenerator}
    \|\xi_B\| \leq G_{u_\alpha} \xi_{u_\alpha} + c_{u_\alpha},
\end{equation}
where the generator set in \eqref{eq:conicGenerator} can be used directly in the CCG latent space.

\subsection{Reachable Set in State-Control Space}
In order to write the distinguishable space, we require encoding the relationship between $x$ and $u$ directly by lifting the set description to a higher dimension. For an easier notation, we present the steps for an LTI to remove the need for the dependence of the scheduling vector, which can be readily generalized once we take the convex hull of all such sets. We start by writing

\begin{equation}
  x_N = A^N x_0 + \sum_{k=0}^{N-1} A^{N-k-1} B u_k
\end{equation}
for which, if we have the initial condition and input described by CCGs of the form $\mathcal{X}_0 = (G_x, c_x, A_x, b_x, \xi_x)$ and $\mathcal{U} = (G_{u}, c_{u}, A_{u}, b_{u}, \xi_{u})$, will result in 
\begin{align}
  \label{eq:vector_sum}
  \mathcal{X}_N &= A^N (G_x \, \xi_x + c_x) +  \mathcal{C}_N  (G_u \, \xi_u + c_u)\\
  &= \begin{bmatrix}
        A^N G_x & \mathcal{C}_N G_u
      \end{bmatrix}
      \begin{bmatrix}
        \xi_x \\ \xi_u
      \end{bmatrix} + A^N c_x + \mathcal{C}_N c_u, \nonumber
\end{align}
meaning
\[
\resizebox{\hsize}{!}{$
\mathcal{X}_N = \left(
\begin{bmatrix} A^N G_x & \mathcal{C}_N G_u \\ 0 & G_u \end{bmatrix},
\begin{bmatrix} A^N c_x + \mathcal{C}_N c_u \\ c_u \end{bmatrix},
\begin{bmatrix} A_x & 0 \\ 0 & A_u \end{bmatrix},
\begin{bmatrix} b_x \\ b_u \end{bmatrix}
\right).$}
\]
In the LPV case, the linear map of the matrix power applied to the initial set $A^N \mathcal{X}_0$ will be replaced by a $N$-th product of all dynamics matrices for each time step $\mathcal{A}^N$. This formulation provides sufficient flexibility to accommodate additional constraints, such as those involving the $\ell_2$-norm, which frequently arise in the context of radial measures or energy-based constraints. 

\section{Set Separation based on Singular Value Decomposition} 
\label{sec:my_svd}

In the previous section, we proposed to write the constraints for the fault-tolerant controller as the condition for distinguishability and resorted to its representation through CCGs. However, the projection part to select the boundary still increases the computing time. In this section, we proposed a suboptimal approach of using the direction of maximum control authority to perform the separation of the nominal and faulty reachable sets. The loss of optimality arises from the fact that the maximum controllable direction can also amplify the effect of the uncertainties. 

Formally, we are searching for a direction such that
\[
  \max\limits_{\|\mathbf{u}\|^2_2 \leq 1} \| x_N^{(1)} - x_N^{(2)} \|^2_2
\]

In order to avoid the aforementioned problem, we will introduce a change of coordinates to account for non-symmetric sets using the inverse of the shape matrix. Since the generator space is typically of higher dimension in comparison with the state-control space, we can compute an over-approximation of $G$ by a square generator matrix $\Phi^{(j)} \in \mathbb{R}^{n_x \times n_x}$ through order reduction for the $j$-th mode. We can perform the transformation and solve the related problem
\begin{align}
    &\max\limits_{\|\mathbf{u}\|^2_2 \leq 1} \|(\Phi^{(1)})^{-1} x_N^{(1)} - (\Phi^{(2)})^{-1} x_N^{(2)}\|^2_2\\
    \iff &\max\limits_{\|\mathbf{u}\|^2_2 \leq 1} \left\| (\Phi^{(1)})^{-1} (\Phi^{(1)} x_0^{(1)} + \mathcal{C}^{(1)}_N\mathbf{u}) - \right.\nonumber\\
    &\qquad \qquad \qquad \left. (\Phi^{(2)})^{-1}(\Phi^{(2)} x_0^{(2)} + \mathcal{C}^{(2)}_N\mathbf{u}) \right\|^2_2
\end{align}
which since both models will have the same initial state value can be simplified further to
\[
  \max\limits_{||\mathbf{u}||^2_2 \leq 1} || ((\Phi^{(1)})^{-1} C^{(1)}_N - (\Phi^{(2)})^{-1} C^{(2)}_N)\mathbf{u} ||^2_2.
\]

Let $V = (\Phi^{(1)})^{-1}C^{(1)}_N - (\Phi^{(2)})^{-1} \mathcal{C}^{(2)}_N$, then the problem corresponds to the maximization of a square function given as

\[
  \max\limits_{||\mathbf{u}||^2_2 \leq 1} || V \mathbf{u} ||^2_2 = \max\limits_{||\mathbf{u}||^2_2 \leq 1} \mathbf{u}^\intercal V^\intercal V \mathbf{u}
\]

By the spectral theorem, we know that a symmetric matrix has an orthonormal basis given by its eigenvectors. We can decompose $V^\intercal V$ as $Q \Lambda Q^\intercal$, where $Q$ is an orthogonal matrix whose columns are the eigenvectors of $V^\intercal V$, and $\Lambda$ is a diagonal matrix with the eigenvalues. 
Since $Q$ is an orthonormal eigen-basis, we can express any vector as $\mathbf{u} = \sum_{i=1}^{n} \alpha_i q_i$, where $\alpha_i$ are the weights, and $q_i$ are the eigenvectors of $V$. The unit norm for $\mathbf{u}$ can be simplified to the sum of the squares of the weights is one, leading to

\begin{align}
  \max\limits_{\sum_{i=1}^n \alpha_i^2 = 1} \mathbf{u}^\intercal V^\intercal V \mathbf{u} &= \max\limits_{\sum_{i=1}^n \alpha_i^2 = 1} \sum_{i=1}^{n} \sum_{j=1}^{n} \alpha_i \alpha_j q_i^\intercal V^\intercal V q_j
\end{align}
which since the eigenvectors form an orthogonal space, the inner product of two distinct eigenvectors is zero, leading to the expression being maximized to be

\[
  \sum_{i=1}^{n} \alpha_i^2 q_i^\intercal V^\intercal V q_i = \sum_{i=1}^{n} \alpha_i^2 \lambda_i q_i^\intercal q_i = \sum_{i=1}^{n} \alpha_i^2 \lambda_i,
\]
where we used the facts $V^\intercal V q_i = \lambda_i q_i$ and that the product of an orthonormal vector with itself is the unity.

Therefore, we are maximizing the weighted sum of the eigenvalues $\lambda_i$ under a unit norm constraint on the factors. The solution corresponds to set weight one on the largest eigenvalue. The direction can be retrieved from the SVD of $V$, where the right singular vector associated with the largest singular value defines the direction of maximal separation. For LPV systems, this procedure must be performed iteratively since the actuation can enlarge the reachable set due to the uncertain components and create overlap. Accordingly, the separation direction and required magnitude are recomputed at each iteration until the process converges within a tolerance.

\section{Simulations}
To test our approach, we start with an LPV model for a hovering drone, 
\[
\begin{aligned}
\dot{p} &= v,  \quad m \dot{v} = R(\phi,\theta,\psi)
\begin{bmatrix}
0 \\ 0 \\ T
\end{bmatrix}
-
\begin{bmatrix}
0 \\ 0 \\ mg
\end{bmatrix}, \\
\dot{R} &= RS(\omega), \quad J \dot{\omega} = \tau - S(\omega) \times (J\omega),
\end{aligned}
\]
where $p$ is the position, $v$ is the velocity of a $m=1\mathrm{Kg}$ quadrotor. The rotation matrix $R$ accounts for the attitude, $g=9.81 \mathrm{m/s^2}$, $\omega$ is the angular velcity, $S(\omega)$ is the skew matrix and $J$ is the inertia matrix with $I_{\phi} = 0.02\mathrm{Kg\cdot m^2},\; I_{\theta} = 0.02\mathrm{Kg\cdot m^2}$, and sampling time $T_d = 0.2\mathrm{s}$. The uncertainty regarding the initial state $\mathcal{X}_0 = \big\{ x \in \mathbb{R}^{10} \;\big|\; \|x\|_{\infty} \le 0.1 \big\}$, the disturbances are contained in $\mathcal{W} = \big\{ w \in \mathbb{R}^{10} \;\big|\; \|w-T_d \, g \, \mathrm{e}_6\|_{2} \le 0.1 \big\}$, where $\mathrm{e}_6$ is the canonical vector with 1 in the 6th entry, and the scheduling vector assumes a maximum error of the yaw angle of $20^ \circ$. 

\begin{figure}[h!]
    \centering
    \includegraphics[width=0.45\textwidth]{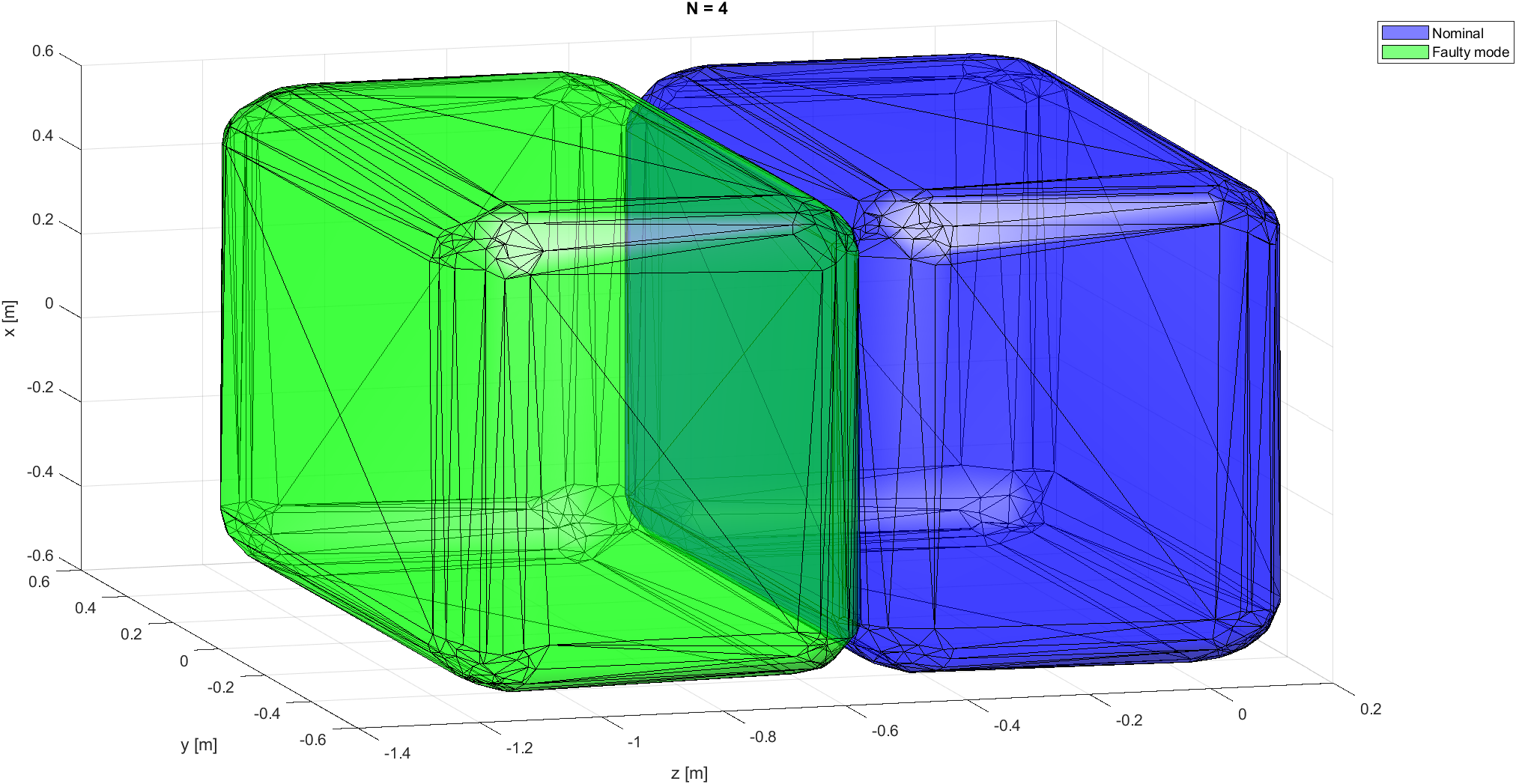} 
    \caption{Reachable sets using the CCG propagation for a horizon of $N=4$ projected on the position coordinates.}
    \label{fig:ccg_sim}
\end{figure}

After a forward Euler discretization and removing the altitude components, we get the LPV:

{\small
\begin{equation*}
A(\theta) = I_{10} + T_d 
\begin{bmatrix}
0 & 1 & 0 & 0 & 0 & 0 & 0 & 0 & 0 & 0 \\
-g\theta_1 & 0 & 0 & 0 & 0 & 0 & 0 & 0 & 0 & 0 \\
0 & 0 & 0 & 1 & 0 & 0 & 0 & 0 & 0 & 0 \\
0 & 0 & -g\theta_2 & 0 & 0 & 0 & 0 & 0 & 0 & 0 \\
0 & 0 & 0 & 0 & 0 & 1 & 0 & 0 & 0 & 0 \\
0 & 0 & 0 & 0 & 0 & 0 & 0 & 0 & 0 & 0 \\
0 & 0 & 0 & 0 & 0 & 0 & 0 & 1 & 0 & 0 \\
0 & 0 & \tfrac{1}{m}\theta_2 & 0 & 0 & 0 & 0 & 0 & 0 & 0 \\
0 & 0 & 0 & 0 & 0 & 0 & 0 & 0 & 0 & 1 \\
-\tfrac{1}{m}\theta_1 & 0 & 0 & 0 & 0 & 0 & 0 & 0 & 0 & 0
\end{bmatrix}
\end{equation*}

\begin{equation*}
B^{(i)}(\theta) = T_d
\begin{bmatrix}
0 & 0 & 0 \\
\tfrac{1}{I_\phi} & 0 & 0 \\
0 & 0 & 0 \\
0 & \tfrac{1}{I_\theta} & 0 \\
0 & 0 & 0 \\
0 & 0 & \tfrac{f^{(i)}}{m}\cos(\theta_1)\cos(\theta_2) \\
0 & 0 & 0 \\
0 & 0 & \tfrac{1}{m}\sin(\theta_1) \\
0 & 0 & 0 \\
0 & 0 & -\tfrac{1}{m}\sin(\theta_2)
\end{bmatrix}
\end{equation*}
}

\begin{equation*}
C =
\begin{bmatrix}
0 & 0 & 0 & 0 & 1 & 0 & 0 & 0 & 0 & 0 \\[4pt]
0 & 0 & 0 & 0 & 0 & 0 & 1 & 0 & 0 & 0 \\[4pt]
0 & 0 & 0 & 0 & 0 & 0 & 0 & 0 & 1 & 0
\end{bmatrix},
\end{equation*}
where $B^{(i)}$ translates the two possible modes with $f^{(1)} = 1$ and $f^{(2)} = 0.5$. The control actions for the roll and pitch torque were assumed in $[0,1]$ and the thrust in $[0,2]$. Using the CCG approach with the cost function
\[\min f(u) = u_4^2 + \sum_{k=1}^{3} \left[ u_k^2 + \big(u_{k+1} - u_k\big)^2 \right] 
\]
we obtain $
\mathbf{u}^\star = 
\begin{bmatrix}
0,  0, 1.537 ,  0,  0,  0.965 ,  0,  0, 0.369 ,  0,  0,  0.052
\end{bmatrix}^\intercal
$
with the reachable sets being depicted in Figure \ref{fig:ccg_sim}.

\begin{figure}[h!]
    \centering
    \includegraphics[width=0.45\textwidth]{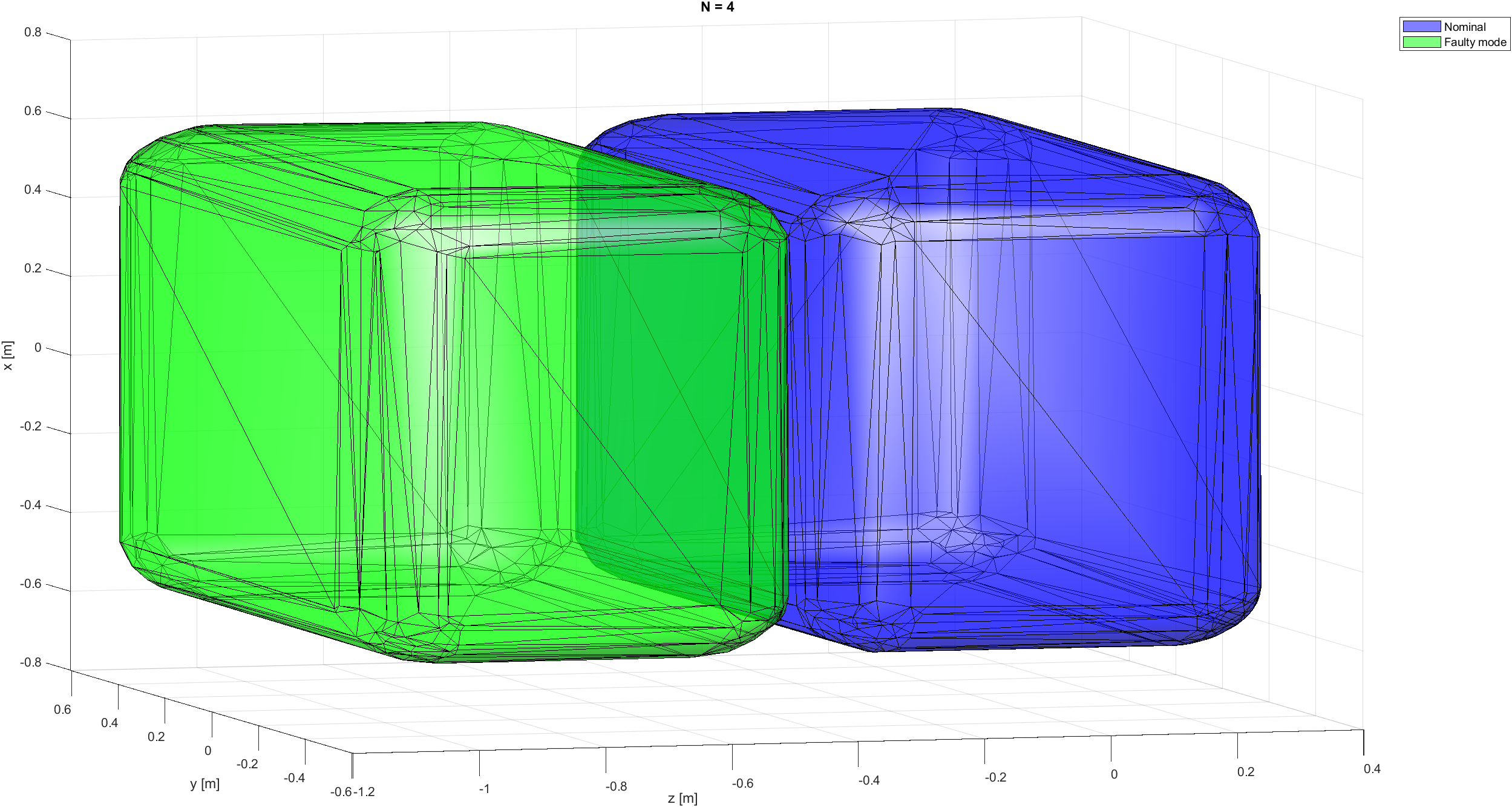} 
    \caption{Reachable sets using the SVD method for a horizon of $N=4$ projected on the position coordinates.}
    \label{fig:svd_sim}
\end{figure}

For the SVD based approach, the computed solution is more inefficient with $
\mathbf{u}^\star = 
\begin{bmatrix}
0,  0, 1.6036 ,  0,  0,  1.0690 ,  0,  0, 0.5345 ,  0,  0,  0
\end{bmatrix}^\intercal$ but the method still allows for an active fault detection as can be seen in Figure \ref{fig:svd_sim}.

In order to compare against the state-of-the-art in \cite{Set-Based-afd}, we consider a model for a vehicle with decoupled axis and unknown coefficients of friction linear with velocity while driving in the ground. In this scenario, the parameters are $\Theta_x = [0.5, 0.8]$, $\Theta_u = [0.7, 1]$, $T_s = 0.5s$, and the maximum drag $D=0.6$
where $\Theta$ is the scheduling vector for the states and control actions. The model matrices are 
\[
A_1(\theta)
=
\begin{bmatrix}
1 - D\,\theta\,T_s & 0 \\
0 & 1 - D\,\theta\,T_s
\end{bmatrix}
B_1(\theta) = 
\begin{bmatrix}
T_s \theta & 0 \\
0 & T_s \theta
\end{bmatrix}
\]

\[
C_1 = I_2 = C2 \quad A_2(\theta) = A_1(\theta) \quad
B_1(\theta) = 
\begin{bmatrix}
0.8T_s \theta & 0 \\
0 & 0.4T_s \theta
\end{bmatrix}.
\]
The sets $\mathcal{X}$, $\mathcal{V}$ and $\mathcal{F}$ all correspond to an $\ell_2$ norm ball of radius 0.1 and a cost function of 

\[\min f(u) = u_3^2 + \sum_{k=1}^{2} \left[ u_k^2 + \big(u_{k+1} - u_k\big)^2 \right] 
\]

\begin{table}[h]
\centering
\caption{Comparison between Our Algorithm and H-rep Algorithm}
\label{tab:comparison}
\begin{tabular}{lcc}
\hline
 & Proposed Algorithm & H-rep Algorithm \cite{Set-Based-afd}\\
\hline
Time & 53 sec & 387 sec \\
Cost & 11.5632 & 19.9649 \\
\hline
\end{tabular}
\end{table}

\begin{figure}[h!]
    \centering
    \includegraphics[width=0.45\textwidth]{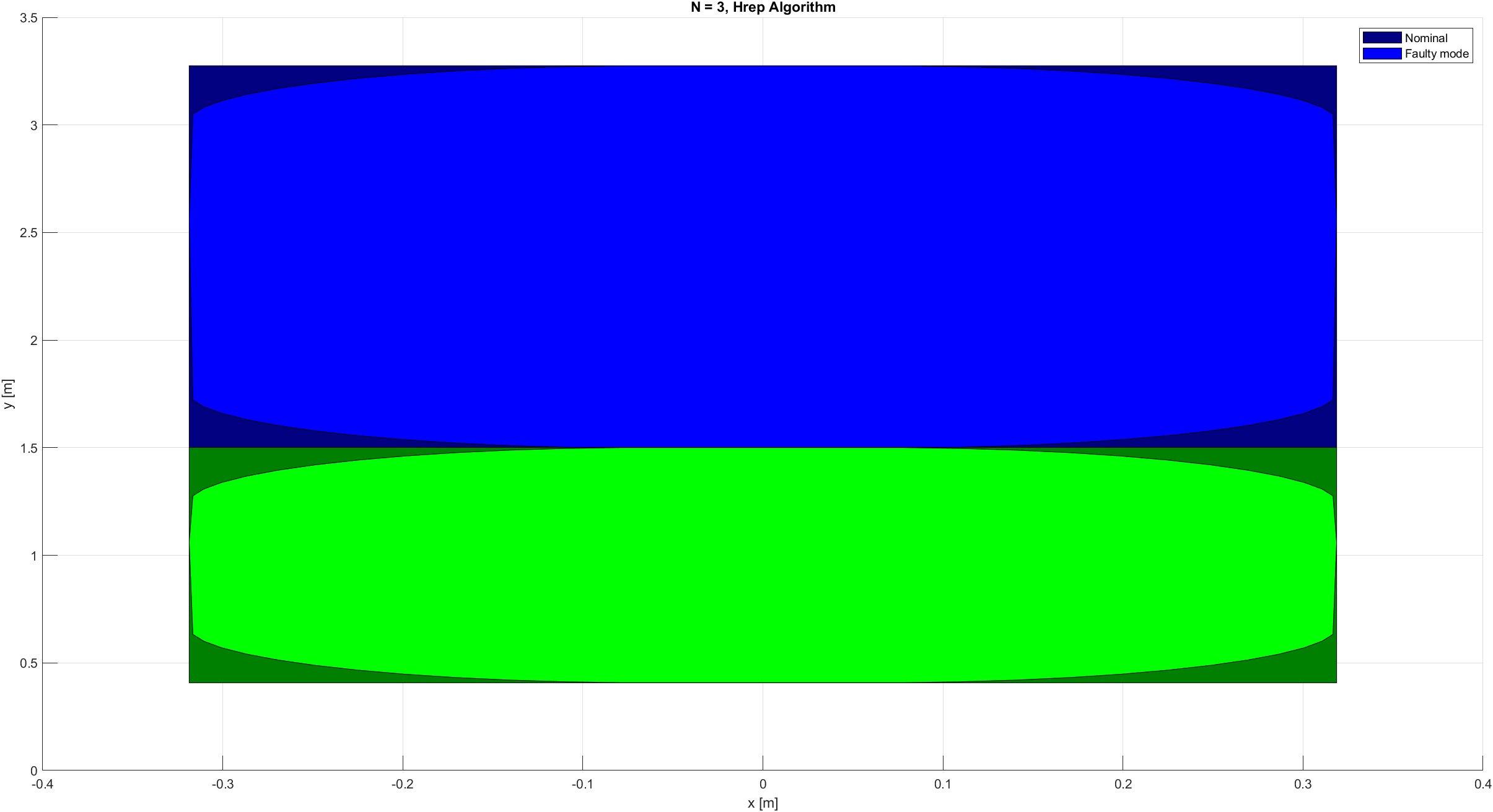} 
    \caption{Reachable set propagation using the method in \cite{Set-Based-afd} that uses $H$-rep polytopes for a horizon of $N=3$.}
    \label{fig:automatica_comp_1}
\end{figure}

Using the CCG method resulted in $\mathbf{u}^\star = 
\begin{bmatrix}
0,  2.7253, 0, 1.7442, 0, 1.0458
\end{bmatrix}^\intercal$ for the proposed algorithm with a cost of $11.56$ whereas the method in \cite{Set-Based-afd} resulted in $
\mathbf{u}^\star = 
\begin{bmatrix}
0,  1.3778, 0, 2.2940, 0, 3.5844
\end{bmatrix}^\intercal$ with a cost of 19.96 as shown in Table \ref{tab:comparison}. Since vertex enumeration in the \cite{Set-Based-afd} is computationally expensive, we employed a ray-shooting method to sample the boundary of the set. For this approach, the number of points required is typically on the order of $2m + 2^m$ for a set in $\R^m$ with the control-action space with $m = 18$, which would require 262,180 points to accurately recover the boundary. Sampling the full range at this resolution is computationally prohibitive, so we limited the number of points to 2,000 which still takes 387 s. In comparison, the proposed method took 53 s because it only samples the boundary at the final step in a 6-dimensional space (we used 484 rays). Moreover, since in \cite{Set-Based-afd} only allows polytopes, improving the accuracy would result in a larger dimension with the additional conservatism shown in the dark regions around the set in Figure \ref{fig:automatica_comp_1}.

\begin{figure}[h!]
    \centering
    \includegraphics[width=0.45\textwidth]{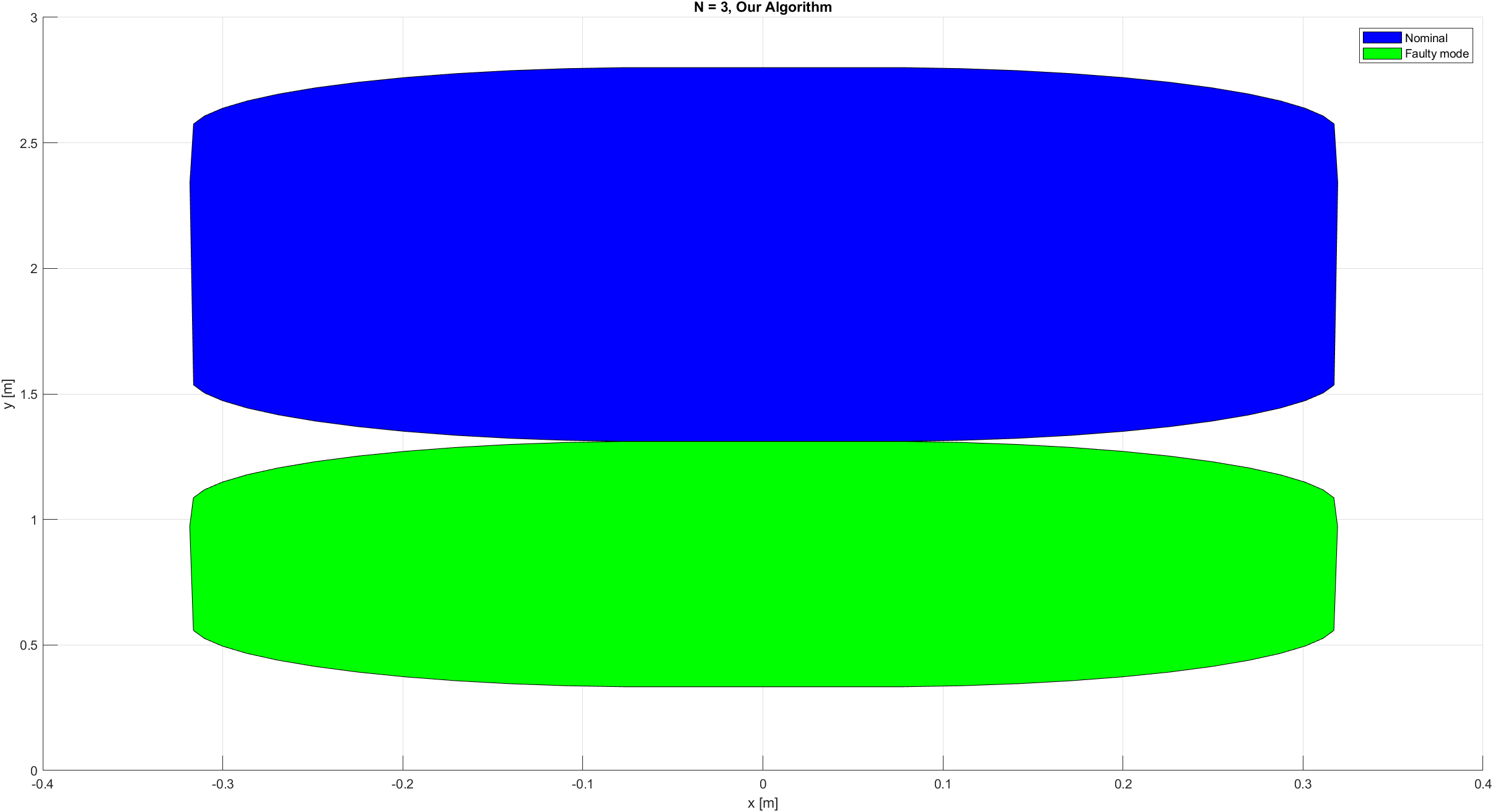} 
    \caption{Reachable set propagation using the proposed CCG method for a horizon of $N=3$.}
    \label{fig:our_comp_1}
\end{figure}

\section{Conclusion}

While the approaches presented in this document provide a solid foundation for fault detection and isolation, there are areas for future improvement. One key aspect is the optimization of the new CCG operations, as well as a better way to obtain the boundary after intersection as that is the most expensive step in the algorithm. Although the current method to get the boundary of a set is effective, if improved could benefit cases where the MPC is not being used for guidance but also for the lower level controller with a shorter sampling time.

\bibliographystyle{IEEEtran}
\bibliography{bibtex/bib/IEEEexample}

\end{document}